# Chirality Dependent Electromagnetically Induced Transparency Based on Double Semi-Periodic Helix Metastructure

Bo Yan,[1,2,*] Fan Gao,[1,2] Hongfeng Ma,[1,3] Kesong Zhong,[1,2] Bin Lv,[1,2] Naibo Chen,[1,2] Pinggen Cai,[1,2] Ziran Ye,[1,2] Yun Li,[1,2] Chenghua Sui,[1,2] Tao Xu,[1,2] Chenghua Ma,[1,2] Qiang Lin[1,2]

[1] *Department of Physics, Zhejiang University of Technology, Hangzhou 310023, China*
[2] *Collaborative Innovation Center for Bio-Med Physics Information Technology of ZJUT, Zhejiang University of Technology, Hangzhou 310023, China*
[3] *Laboratoire Hubert Curien, UMR CNRS 5516/UJM/University of Lyon, 42000 Saint-Etienne, France*
*Corresponding author: boyan@zjut.edu.cn

A chiral metastructure composed of spatial separated double semi-periodic helices is proposed and investigated theoretically and experimentally in this work. Chirality dependent electromagnetically induced transparency (EIT) and slow light effect in microwave region are observed from numerical parameter study, while experimental results from the 3D printing sample yields good agreement with the theoretical findings. The studied EIT phenomenon arises as a result of destructive interference by coupled resonances and the proposed chiral metastructure can be applied in polarization communication, pump-probe characterization and quantum computing areas, etc.

**Keywords:** Materials: Chiral media; Materials: Metamaterials; Physical optics: Polarization.

Optical chirality is currently attracting growing attention in both fundamental theory and potential applications [1-5]. Tremendous efforts have been devoted to the studies and open a new realm in spin optics, where the photonic spins (denoted by $\sigma+ = 1$ and $\sigma-=1$) are corresponding to right-handed circular polarization (RCP) and left-handed circular polarization (LCP) respectively [6-11]. Moreover, the spin-dependent spatial intensity difference can be further developed as an emerging technique for detecting Stokes vectors, hence, enabling complete Stoke vector imaging [12-14]. Observation of a spin-dependent phenomenon in the time domain is possible via index circular polarization selective media. Nonetheless, finding such materials in nature is not an easy task. Fortunately, the effect of electromagnetic induced transparency (EIT) provides a specific route for constructing such materials to manipulate light pulses.

EIT is usually referred to a quantum destructive interference effect observed in ultra-cold vapors of metal atoms with a narrow transmission window and unusual dispersive properties [15-19]. Analogues of EIT effects are not restricted to quantum systems and have been proposed in classical systems like mechanical oscillators and metamaterials [20-25]. By adjusting the geometry parameters of the EIT metamaterials, the operating regions and dispersive properties can be manipulated. To date, previous works mainly focused on metamaterials and metastructures with linear polarization selectivity, the EIT-like phenomenon with circular polarization selectivity has been seldom involved [26,27]. Here in this letter, we propose a new type of metastructure composed of two spatial separated semi-periodic helices, from which the numerical parameter study presents a chirality dependent EIT phenomenon and slow light effect. Furthermore, we fabricate the metastructure sample with 3D printing technique and the experimental findings agree well with the theoretical work. Although studied for a specific system in microwave region, the mechanism of the chirality dependent EIT phenomenon is general and could be extended to the Terahertz and infrared-visible region.

The unite cell of the designed metastructure is illustrated in Fig. 1, which has two Ag semi-periodic helices with fiber diameter $d$, major radius $R$, axial pitch $h$, lattice constant $p$, axial separation $\Delta h$, thickness of substrate $h_S$ and height of cylinder $H_C$. The identical half helices are twined on the PTFE pillars (PTFE Arlon AD 300 plastics, $\epsilon = 3$). The transmission spectra of the proposed metastructure ($h = 11$ mm, $d = 1.4$ mm, $D = 6$ mm, $p = 14$ mm, $\Delta h = 2.83$ mm, $h_S = 2$mm and $H_C = 28$mm) for LCP and RCP are depicted in Fig. 2(a). The numerical calculation is carried out by the commercial software package (COMSOL Multiphysics) using finite element method. Clearly, around the resonance of low frequency at 6.165 GHz, a narrow transparency peak and a shallow dip are investigated for LCP and RCP, respectively. Figure 2(c) demonstrates schematic formation of the EIT-like phenomenon. Under strong resonance of single helix, the normal incident LCP wave is mainly reflected that makes ultra-small part of light across the cutting plane and reaches the second helix. Since the second helix is identical to the first one in light propagation path, it strongly oscillates in the same frequency and serves as a dark mode. Apparently, the coupling between helices is mainly through electromagnetic induction. The propagation length from the end of the first helix to the end of the second helix is L = h+Δh = 13.83 mm, at meanwhile the wavelength in the PTFE plastics is λ=28.10 mm at 6.165 GHz. Thus, the synchronous currents [see Fig. 2(d)] are a manifestation of a destructive interference between the symmetric modes that lead to high-amplitude oscillations, which is analogous to mechanical EIT systems with identical oscillators [20,28]. In terms of RCP, the field is the sum of the incident light and the induction coupling response. Considerable electromagnetic waves transmit the first helix and induce a weak resonance in the second helix. Comparing to the direct excitation by RCP, the induced currents by coupling propagates inversely for the phase delay. As a result, a transmission dip emerges due to the absorption.

Around the high-frequency at 10.29 GHz, EIT-like profiles are observed in both LCP and RCP [see Fig. 2(a)]. In this situation, the currents in the two helices move in counter-propagating direction. Similarly, the high-frequency response of the metastructure are the result of destructive interference by the asymmetry modes (propagation length L = 13.83 mm, λ=16.833 mm at 10.29 GHz). In the case of normal illumination by LCP, the transmitted waves no longer eliminate the induction currents by coupling in helix 2. However, things would be different and more complex when the propagation length is close to λ/2.

The metastructure sample is fabricated to verify the simulation results. The Nylon substrate and cylinder arrays are prepared with SLS Nylon 3D printing technique. Double semi-periodic helices of Ag are then twined on the plastic pillars with UV-adhesives as showing in the insert of Fig.3. The transmission spectra present EIT-like profiles with sharp transparency peak and shallow dip around 6.15 GHz for LCP and RCP respectively [see Fig.3], while for high frequency of 10.29 GHz corresponding EIT peaks also emerge in both LCP and RCP curves, which yield good agreement with the simulation results. The lower intensity and broadened transparency peaks are due to the dielectric constant changing of PTFE plastics and the air. Furthermore, the errors during Ag-helices assembling also contribute to these differences.

In atomic EIT systems, the coupling between two energy levels depends strongly on the pump beam intensity. In order to observe a narrow dip in the absorption power of the probe laser, the pumping laser should be intense enough so that the Rabi frequency is larger than all damping rates. Moreover, an increment of the coupling coefficient would lead to the apparition of two peaks in the absorption power of the probe laser [29]. In contrast, the separated helices systems present a more complex phenomenon when the axial separations are changed [see Fig. 4], where the coupling between bright and dark modes is determined by the spatial separation. A decrement in axial separation does not lead to an apparition of two dips, which, however, brings about a Fano-like profile in LCP transmission spectra accompany with a blue shift of the peak [30]. Interestingly, a further decrement results in a broad stop band. This phenomenon is somewhat similar to single or two pitches helix metamaterials, where the Bragg resonance is believed to play a key role in the broadband response [31-33]. The transmission profile for a little increment (Δh = 3.01 mm) in axial separation shows no differences with the sharp peak situation (Δh =2.96 mm) apart from the peak magnitude. This is rather similar to the atomic EIT systems suffer from strong absorptions: as the coupling intensity decreases, the transmission becomes weaker and will stop finally. To our surprise, the further increments in axial separation present sharp transmission profiles. These observations indicate that the change in axial separation affects the coupling, the current distribution and the wave interference.

Another important evidence of the EIT-like phenomenon is the dispersion on resonance. To this end, we retrieved the effective index ($n_p$) with the phase information of the transmittance function. Then, the group index [see Fig. 5] can be obtained by

$$n_g = n_p(\omega) + \omega \frac{dn_p}{d\omega}. \quad (1)$$

As the Kramers-Kronig relation suggests, strong dispersion appears in the case of LCP. To verify the slow light effect, we use finite difference time domain [34] to investigate the time evolution of a Gaussian-shaped pulse centered at 6.161 GHz with half-maximum at full-width of 80ns. The time evolution of electric field intensity for LCP and RCP are shown in insert map of Fig. 5. One can see that the pulse intensity profile of LCP peaks at 178092 ps with a time delay of $\tau_L$ = 18092 ps. The corresponding group index is estimated as $c_0 * \tau_L / L_{SO} \approx 96.9$, where $c_0$ is light speed in vacuum and $L_{SO}$ = 56mm is the length from source to observer, which agrees in order of magnitude with the retrieval method. The number can be larger for the pulse with narrower width of frequency. In contrast to LCP, the RCP pulse peaks at 160329 ps and shows $\tau_R$ = 329 ps delay. This delay is mainly due to the transportation from light source to observer. The chiral dependent strong EIT-like interaction leads to 17 ns time delay between LCP and RCP that opens a new route to engineering circular polarization directly and provides a new way to understand light intrinsic spin [35].

In conclusion, we have proposed a spatial separated double semi-periodic helix metastructure and investigated theoretically and experimentally. Novel chirality-dependent EIT phenomenon in microwave region is observed with numerical calculation and further proved by the measurement of the 3D printing sample. This effect results from the coherent process modified by the chiral nature of individual helix. Under resonance, LCP and RCP electromagnetic pulses transmit the structure with different group velocities, thus occurs the slow light effect. This work provides a new route for studying spin-dependent effect as well as the chiral EIT-like phenomenon in time domain and paves the way for its applications in polarization communication and quantum computing regions.

**Funding.** National Nature Science Foundation of China (NSFC) (61705197, 11604295), Zhejiang Provincial Nature Science Foundation of China (LQ17C100002)

**Acknowledgment**. We thank Prof. Yi Shi of Nanjing University and Prof. Gaoxiang Ye of Zhejiang University for their helpful discussions on this work.

**References**
1. P. Lodahl, S. Mahmoodian, S. Stobbe, A. Rauschenbeutel, P. Schneeweiss, J. Volz, H. Pichler, and P. Zoller, Nature **541**, 473 (2017).
2. I. M. Mirza, J. G. Hoskins, and J. C. Schotland, Phys. Rev. A **96** (2017).
3. C. Gonzalez-Ballestero, A. Gonzalez-Tudela, F. J. Garcia-Vidal, and E. Moreno, Phys Rev B **92** (2015).
4. I. M. Mirza and J. C. Schotland, J. Opt. Soc. Am. B **35**, 1149 (2018).
5. I. M. Mirza and J. C. Schotland, Phys. Rev. A **94** (2016).
6. K. Y. Bliokh, F. J. Rodriguez-Fortuno, F. Nori, and A. V. Zayats, Nat. Photonics **9**, 796 (2015).
7. C. P. Jisha and A. Alberucci, Opt. Lett. **42**, 419 (2017).
8. A. V. Kildishev, A. Boltasseva, and V. M. Shalaev, Science **339** (2013).
9. Y. Li, Y. C. Liu, X. H. Ling, X. N. Yi, X. X. Zhou, Y. G. Ke, H. L. Luo, S. C. Wen, and D. Y. Fan, Opt. Express **23**, 1767 (2015).
10. N. Shitrit, I. Yulevich, E. Maguid, D. Ozeri, D. Veksler, V. Kleiner, and E. Hasman, Science **340**, 724 (2013).
11. X. B. Yin, Z. L. Ye, J. Rho, Y. Wang, and X. Zhang, Science **339**, 1405 (2013).
12. D. Maluenda, R. Martinez-Herrero, I. Juvells, and A. Carnicer, Opt. Express **22**, 6859 (2014).
13. K. A. Bachman, J. J. Peltzer, P. D. Flammer, T. E. Furtak, R. T. Collins, and R. E. Hollingsworth, Opt. Express **20**, 1308 (2012).
14. B. Yan, K. S. Zhong, H. F. Ma, Y. Li, C. H. Sui, J. Z. Wang, and Y. Shi, Opt. Commun. **383**, 57 (2017).
15. A. H. Safavi-Naeini, T. P. M. Alegre, J. Chan, M. Eichenfield, M. Winger, Q. Lin, J. T. Hill, D. E. Chang, and O. Painter, Nature **472**, 69 (2011).
16. M. F. Limonov, M. V. Rybin, A. N. Poddubny, and Y. S. Kivshar, Nat. Photonics **11**, 543 (2017).
17. K. J. Boller, A. Imamoglu, and S. E. Harris, Phys. Rev. Lett. **66**, 2593 (1991).
18. L. V. Hau, S. E. Harris, Z. Dutton, and C. H. Behroozi, Nature **397**, 594 (1999).
19. C. Liu, Z. Dutton, C. H. Behroozi, and L. V. Hau, Nature **409**, 490 (2001).
20. P. Tassin, L. Zhang, R. K. Zhao, A. Jain, T. Koschny, and C. M. Soukoulis, Phys. Rev. Lett. **109** (2012).
21. P. Pitchappa, M. Manjappa, C. P. Ho, R. Singh, N. Singh, and C. Lee, Adv. Opt. Mater. **4**, 541 (2016).
22. S. G. Lee, S. Y. Jung, H. S. Kim, S. Lee, and J. M. Park, Opt. Lett. **40**, 4241 (2015).
23. J. L. Liu, H. J. Yang, C. Wang, K. Xu, and J. H. Xiao, Sci. Rep.-Uk **6** (2016).
24. G. X. Wang, H. Lu, and X. M. Liu, Opt. Express **20**, 20902 (2012).
25. P. Tassin, L. Zhang, T. Koschny, E. N. Economou, and C. M. Soukoulis, Phys. Rev. Lett. **102** (2009).
26. H. Lin, D. Yang, S. Han, Y. J. Liu, and H. L. Yang, Opt. Express **24**, 30068 (2016).
27. L. Zhu, F. Y. Meng, L. Dong, J. H. Fu, F. Zhang, and Q. Wu, Opt. Express **21**, 32099 (2013).
28. N. Papasimakis, V. A. Fedotov, N. I. Zheludev, and S. L. Prosvirnin, Phys. Rev. Lett. **101** (2008).
29. C. L. G. Alzar, M. A. G. Martinez, and P. Nussenzveig, Am. J. Phys. **70**, 37 (2002).
30. B. Luk'yanchuk, N. I. Zheludev, S. A. Maier, N. J. Halas, P. Nordlander, H. Giessen, and C. T. Chong, Nat. Mater. **9**, 707 (2010).
31. J. K. Gansel, M. Thiel, M. S. Rill, M. Decker, K. Bade, V. Saile, G. von Freymann, S. Linden, and M. Wegener, Science **325**, 1513 (2009).
32. J. K. Gansel, M. Wegener, S. Burger, and S. Linden, Opt. Express **18**, 1059 (2010).
33. Y. Yu, Z. Y. Yang, S. X. Li, and M. Zhao, Opt. Express **19**, 10886 (2011).
34. A. F. Oskooi, D. Roundy, M. Ibanescu, P. Bermel, J. D. Joannopoulos, and S. G. Johnson, Comput. Phys. Commun. **181**, 687 (2010).
35. B. Wu, J. F. Hulbert, E. J. Lunt, K. Hurd, A. R. Hawkins, and H. Schmidt, Nat. Photonics **4**, 776 (2010).

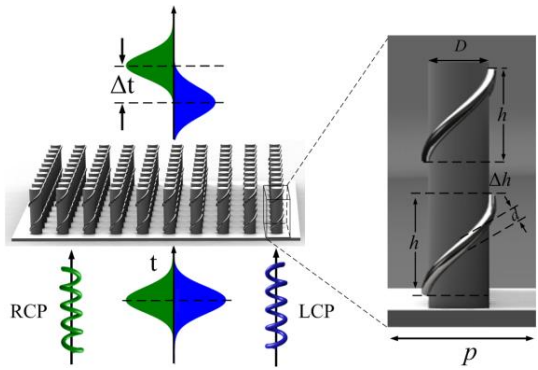

Fig.1. Schematic of the studied chiral metastructure and the unite cell. Two identical semi-periodic helices of silver are embedded in a dielectric medium. The green and blue helices represent for RCP and LCP light, respectively.

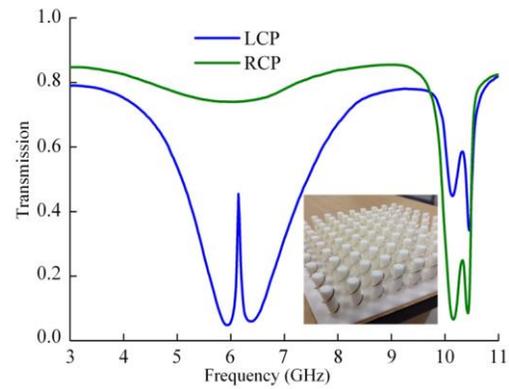

Fig.3. The transmission spectra of the 3D printed 10∗10 sample present obvious EIT-like phenomenon around 6.15 GHz for LCP and RCP. The insert is the photo of the real sample which Ag semi-periodic helices are twined around the PTFE pillar.

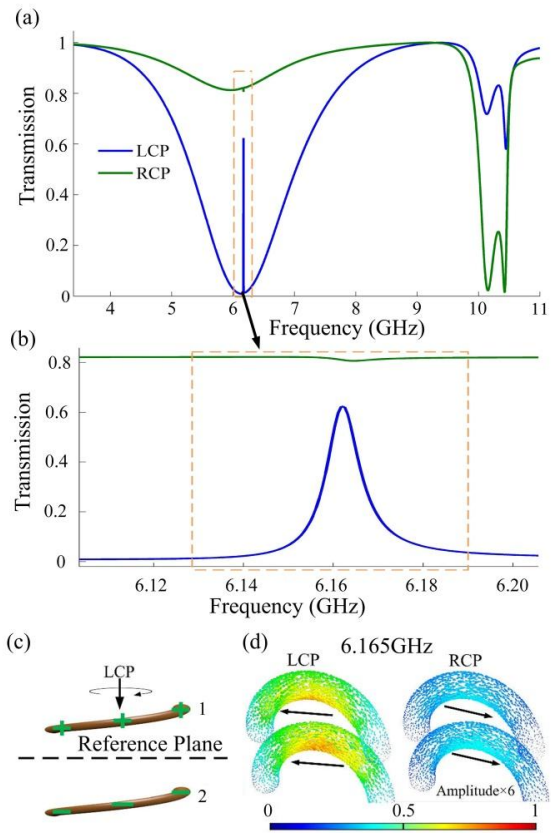

Fig.2. (a) The transmission spectra of the studied chiral metastructure for LCP and RCP light. (b) The transmission profiles around the peak position at a higher magnification. (c) Schematic formation of EIT-like phenomenon. (d) Synchronous current map in the half pitch helices at 6.165GHz under LCP and RCP, respectively.

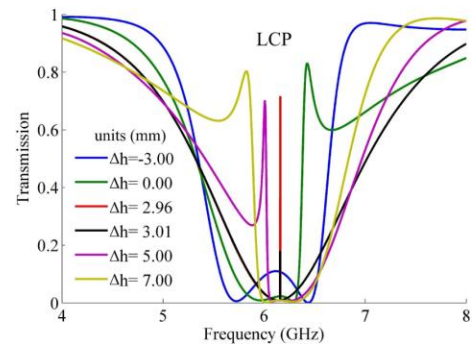

Fig.4. The LCP transmission spectra of the double semi-periodic helix metastructure with different axial separations.

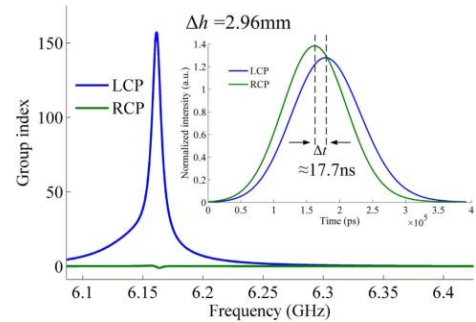

Fig.5. The retrieved group index from the transmission function. The insert map shows the time evolution of a Gaussian-shape pulse for LCP and RCP, respectively.